\documentstyle[12pt]{article}

\begin{document}

\author{Sukratu Barve and T. P. Singh  \\ 
Theoretical Astrophysics Group\\Tata Institute of Fundamental Research\\Homi
Bhabha Road, Bombay 400 005, India}
\title{Are naked singularities forbidden by the second law of thermodynamics?}
\date{}
\maketitle

\begin{abstract}
\noindent By now, many examples of naked singularities in classical general
relativity are known. It may however be that a physical principle over and
above the general theory prevents the occurrence of such singularities in
nature. Assuming the validity of the Weyl curvature hypothesis, we propose
that naked singularities are forbidden by the second law of thermodynamics.
\end{abstract}

\newpage\ 

\noindent The cosmic censorship hypothesis remains one of the most important
unsolved problems in classical general relativity. Put in physical,
non-rigorous, terms the hypothesis states that generic singularities arising
in gravitational collapse are not visible. This hypothesis is crucial for
the validity of black-hole physics and astrophysics. For instance, important
theorems like the area-increase theorem for black-holes assume the validity
of this hypothesis. The violation of the hypothesis implies that visible
singularities can arise in gravitational collapse - these are called naked
singularities.

Over the years, attempts to provide a proof of the censorship hypothesis
have not been successful. On the other hand, many examples of naked
singularities have been found in models of 
gravitational collapse in general relativity.
Most of these examples have been found in studies of spherical collapse.
These include (i) naked singularities in the collapse of inhomogeneous dust
described by the Tolman-Bondi spacetime \cite{tb}, (ii) naked singularities
in the collapse of null dust described by the Vaidya spacetime \cite{v},
(iii) naked singularities in the collapse of perfect fluids \cite{pf}, (iv)
naked singularity at the critical point in the collapse of a massless scalar
field \cite{sf}. There is also some evidence (though not incontrovertible)
of naked singularity formation in the non-spherical collapse of dust \cite
{st}. While these examples are not sufficiently generic as to already
invalidate the censorship hypothesis, their increasing number seriously
suggests the possibility that the hypothesis may not be true after all. If
so, black holes and naked singularities must be considered at par, from the
point of view of classical general relativity.

The occurrence of a naked singularity represents the breakdown of
predictability, because completely arbitrary data could be received by an
observer who is to the future of the singularity. One might be compelled to
consider modifying general relativity, so as to avoid naked singularity
formation, and hence preserve predictability. However, as we discuss below,
there appears to be a less drastic way out, which is consistent with
presently known laws of physics, and does not require modification of the
theory.

While general relativity does admit dynamical solutions leading to naked
singularities, we should ask if these solutions are realized in the real
physical world. Could it be that there is a physical principle, over and
above general relativity, which forbids their occurrence in nature? For
comparison, we recall the well-known advanced wave solutions of Maxwell's
electrodynamics, whose occurrence in the real world is forbidden by the
second law of thermodynamics. In a similar spirit, we propose in this essay
that it is the second law which prevents the occurrence of naked
singularities in nature. Naked singularities are seen to be analogous to the
advanced wave solutions in electromagnetism - allowed by the theory, but not
observed in the real world.

Our proposal is intimately connected with the Weyl curvature hypothesis put
forward by Penrose \cite{wch}. As emphasized by Penrose, the entropy of the
universe at its beginning must have been extremely low, in order for there
to be something like the second law. There is very good evidence, however,
that matter itself was in a high entropy, thermal equilibrium state. 
Thus matter by itself could not have been responsible for the overall low
entropy. Hence
the low entropy constraint at the Big Bang must come from the spacetime
geometry - a suitably defined gravitational entropy must have had a
remarkably low value. It is highly plausible that it is the Weyl curvature
tensor which is a measure of gravitational entropy. This is because an
initially Friedmannean universe will have zero Weyl curvature (and hence
zero gravitational entropy). As the universe evolves, irregularities and
clumping in the matter distribution develop; these result in a continual
increase of the Weyl curvature and of gravitational entropy. Hence the
hypothesis is stated as: the Weyl curvature of the universe is zero at the
initial singularity. The second law of thermodynamics is then to be
understood as a consequence of this hypothesis.

If cosmic censorship is assumed to be valid, then the final singularities,
which are a result of gravitational collapse, will be black-holes. These
singularities will have resulted from ever-increasing gravitational
clumping, and typically the Weyl tensor will diverge in the approach to the
black-hole singularity. Hence the structure of the final singularity will be
greatly different from that of the initial singularity.

Our interest is in the implications of the Weyl curvature hypothesis if
cosmic censorship fails. If a naked singularity results as the end state of
gravitational collapse, it classifies as an {\it initial} singularity,
because there are outgoing geodesics which terminate in their past at the
singularity. The validity of the Weyl hypothesis requires that as the naked
singularity is approached in the past along an {\it outgoing}
geodesic, the Weyl
curvature should go to zero \cite{wch2}. This is essential because if
general relativity admits a naked singularity solution in collapse, such a
solution could in principle be indistinguishable from an initial
cosmological singularity, for which case the hypothesis requires Weyl to be
exactly zero.

Since a few analytical examples of naked singularities are known, it is
easily possible to calculate the growth of the Weyl curvature along outgoing
geodesics, as the singularity is approached in the past. Hence one could
check for the validity of the hypothesis. For definiteness, we assume that
it is the Weyl scalar $C\equiv C^{abcd}C_{abcd}$ which has to be
zero at the initial singularity. Here we examine two models of spherical
collapse which give rise to naked singularities. One is the spherical
gravitational collapse of inhomogeneous dust, which is described by the
Tolman-Bondi spacetime, for which the metric in comoving coordinates $%
(t,r,\theta ,\phi )$ is

\begin{equation}
\label{tb}ds^2=dt^2-\frac{R^{\prime 2}}{1+f(r)}dr^2-R^2d\Omega ^2 
\end{equation}
Here, $R(t,r)$ is the area radius at time $t$ of a shell with comoving
coordinate $r$, and $f(r)$ is a free function which is determined by the
initial density and velocity distribution of the dust cloud. The Weyl scalar 
$C$ is equal to 
\begin{equation}
\label{wtb}C(t,r)=\frac{48}{R^4}\left( \frac{m(r)}R-\frac{m^{\prime }(r)}{%
3R^{\prime }}\right) ^2 
\end{equation}
where $m(r)$ is the mass of the cloud interior to $r$. Note that $m(r)$ is
time independent, (since we are dealing with dust) - it is determined by the
initial density distribution. This scalar remains zero at the origin
throughout the nonsingular phase of the evolution, but blows up at the
curvature singularity, where the Kretschmann scalar also diverges.

It is known \cite{tb} that for certain initial density distributions, the
singularity at $R=0,r=0$ resulting in dust collapse is at least locally
naked, and there is a family of null geodesics emanating from the
singularity. We define the tangent along an outgoing geodesic to be $X\equiv
R/r^{\alpha }$ where $\alpha $ is a constant greater than unity,
chosen so that the tangent is well-defined \cite{tb}. It can be shown that
the limiting value of the Weyl scalar as the naked singularity is approached
in the past along an outgoing geodesic is

\begin{equation}
\label{wtb2}C(X_{0},r)=\frac{16\rho _0^2\theta _0^2}{3X_0^6(X_0^{3/2}+\theta
_0)^2}r^{6(1-\alpha )} 
\end{equation}
Here, $\rho_0 $ is the initial central density, $X_0$
is the limiting value of the tangent, and $\theta _0$ is a finite quantity
determined by the initial density and velocity profile in the neighborhood
of the origin. As we easily see by letting $r\rightarrow 0$, the Weyl scalar
diverges at the naked singularity, along an outgoing geodesic. The same is
true along an ingoing geodesic. We note that the divergence of the Weyl
scalar and the occurrence of the naked singularity are both a consequence of
the inhomogeneity of the initial density distribution. In a homogeneous
collapse (the Oppenheimer-Snyder solution) the Weyl is exactly zero inside
the cloud, and the collapse results in the formation of a black-hole.
Also, the scalar is calculated using the Tolman-Bondi metric as the outgoing
rays initially travel through the dust cloud.

As a second example, we consider the self-similar collapse of null dust.
This is described by the Vaidya spacetime, which has the metric

\begin{equation}
\label{va}ds^2=\left( 1-\frac{2m(v)}r\right) dv^2-dr^2-r^2d\Omega ^2 
\end{equation}
Here $v$ is the advanced time coordinate, and $m(v)$ the mass function. For
self-similar collapse, $m(v)=\lambda _0v$, $\lambda _{0}$ being a
constant. This constant represents the rate of infall, $dm(v)/dv$, of the
null dust. It is known that the singularity at $r=0,v=0$ resulting in
collapse is naked for $\lambda _0\leq 1/8$ \cite{v}. The Weyl scalar $C(v,r)$
is equal to $48m^2/r^{6}$. The tangent to an outgoing geodesic which
meets the naked singularity in the past is defined as $X=v/r$. Thus the Weyl
scalar diverges as $\lambda _0^2X_0^2/r^4$ along the outgoing geodesic from
the naked singularity, $X_0$ being the limiting value of the tangent. Once
again the divergence of the Weyl and the occurrence of the naked singularity
are both a consequence of the inhomogeneity of the spacetime.

We see that in both examples, the Weyl scalar diverges as the naked
singularity is approached in the past. This is contrary to the expectations
of the Weyl curvature hypothesis. In fact, since naked singularities in
general arise when the matter distribution is highly inhomogeneous, one
should actually expect the Weyl to {\it diverge} along outgoing geodesics, 
instead
of going to zero. Hence the occurrence of naked singularities in collapse
corresponds to {\it initial} singularities of very high Weyl curvature, and 
equivalently, very high gravitational entropy.

In order for the second law to hold, such initial singularities must be
disallowed. Thus we are presented with two possibilities: (i) The Weyl
curvature hypothesis and the suggested relation between Weyl and
gravitational entropy is not true, and naked singularities occur in the real
world. Or, alternatively, (ii) the Weyl hypothesis is true, and as a
consequence, the second law forbids the occurrence of naked singularities,
because these represent {\it initial} configurations of very high entropy.

Which of these two alternatives is more likely to be the correct one? In our
view, the Weyl hypothesis is an important and perhaps the unique step
towards an understanding of the second law. Hence it should not be
sacrificed. On the other hand, naked singularities are predicted by general
relativity, but apparently not observed in the real world. Hence we propose
that it is thermodynamics, via the Weyl curvature hypothesis, which forbids
the occurrence of naked singularities in nature. This would imply that
somehow nature remarkably  avoids those initial conditions in stellar
collapse which eventually lead to naked singularities.

\end{document}